\documentclass{appolb}
\usepackage{slashed}
\usepackage{graphicx}
\usepackage{color}

\begin{document}
\title{$\eta'$ multiplicity and Witten-Veneziano relation at $T>0$
 }%
\author{D. Klabu\v car$^1$, S. Beni\'c$^1$, D. Horvati\'c$^1$, D. Kekez$^2$
\address{$^1$Physics Department, Faculty of Science, University of Zagreb, Croatia \\
$^2$Rugjer Bo\v skovi\'c Institute, Bijeni\v{c}ka c. 54, Zagreb 10000, Croatia }
}
\maketitle
\begin{abstract}
\noindent
Recent RHIC results on $\eta'$ multiplicity in heavy-ion 
collisions are of great importance because they clearly 
signal a partial restoration of $U_A(1)$ symmetry at 
high temperatures $T$, and thus provide an unambiguous 
signature of the formation of a new state of matter.
Prompted by these experimental results of STAR and PHENIX
collaborations, we discuss and propose the minimal 
generalization of the Witten-Veneziano relation to 
finite $T$.

\end{abstract}
\PACS{11.10.St, 11.10.Wx, 12.38.-t, 24.85.+p}

%
%
\section{Introduction}
\label{Intro}

The most compelling signal for production of a new form of QCD matter
at heavy-ion collider facilities like RHIC and LHC - i.e., strongly 
coupled quark-gluon plasma (sQGP) -
would be a restoration, in hot and/or dense matter, of the
symmetries of the QCD Lagrangian which are broken in the vacuum.
One of them is the [SU$_A$($N_f$) flavor] chiral symmetry. Its
dynamical breaking results in light, (almost-)Goldstone
pseudoscalar ($P$) mesons -- namely the octet
$P = \pi^0,\pi^\pm,K^0, {\bar K}^0, K^\pm,\eta$
(for three light quark flavors, $N_f=3$).
The second one
is the U$_A$(1) symmetry. Its breaking by the non-Abelian axial
Adler-Bell-Jackiw anomaly (`gluon anomaly' for short) makes
the remaining pseudoscalar meson of the light-quark sector, the
$\eta'$, much heavier, preventing its appearance as the ninth
(almost-)Goldstone boson of dynamical chiral symmetry breaking
(DChSB) in QCD.

The first experimental signature of a partial restoration of the U$_A$(1)
symmetry seems to have been found in the $\sqrt{s_{N\!N}} = 200$ GeV central
Au+Au reactions at RHIC.     Namely, Cs\"org\H{o} {\it et al.} 
\cite{Csorgo:2009pa} analyzed the combined PHENIX 
\cite{Adler:2004rq} and STAR \cite{Adams:2004yc} data very robustly,
through six popular models for hadron multiplicities, and found that
at 99.9\% confidence level, the $\eta'$ mass in the vacuum,
$M_{\eta'}=958$ MeV, is reduced by at least 200 MeV inside the fireball.
It is the sign of the disappearing
contribution of the gluon axial anomaly to the $\eta'$ mass,
which would drop to a value readily understood together with
the (flavor-symmetry-broken) octet of
$q\bar q'$ ($q,q'=u,d,s$) pseudoscalar mesons.
This is  the ``return of the prodigal Goldstone boson"
predicted \cite{Kapusta:1995ww}
as a signal of the U$_A$(1) symmetry restoration.

A related theoretical issue, which we want to address here, is 
the status, at $T>0$, of the famous Witten-Veneziano relation 
(WVR) \cite{Witten:1979vv,Veneziano:1979ec}
\begin{equation}
M_{\eta'}^2 + M_\eta^2 - 2 M_K^2 = \frac{6 \, \chi_{\rm Y\!M}}{f_\pi^2} \,
\label{WittenVenez}
\end{equation}
between the $\eta'$, $\eta$ and $K$-meson masses $M_{\eta',\eta,K}$,
pion decay constant $f_\pi$, and Yang-Mills (YM) topological 
susceptibility $\chi_{\rm Y\!M}$.  It is well satisfied at $T=0$ 
for $\chi_{\rm Y\!M}$ obtained by lattice calculations (e.g.,
\cite{Lucini:2004yh,DelDebbio:2004ns,Alles:2004vi,Durr:2006ky}).
Nevertheless, the $T$-dependence of $\chi_{\rm Y\!M}$ is such 
\cite{Horvatic:2007qs} that the straightforward extension of 
Eq. (\ref{WittenVenez}) to $T>0$ \cite{Horvatic:2007qs},
i.e., replacement of all quantities
therein by their respective
$T$-dependent versions $M_{\eta'}(T)$, $M_{\eta}(T)$, $M_K(T)$,
$f_\pi(T)$ {\it and} $\chi_{\rm Y\!M}(T)$, leads to a conflict with
experiment \cite{Csorgo:2009pa}.
Nevertheless, this paper details a mechanism, proposed in
Refs. \cite{Benic:2011fv,Benic:2012eu}
which enables WVR to agree with experiment at $T>0$.

%
%

%
%
%

\section{The relations connecting two theories, QCD and YM}
\label{relationsConnect}

The dependence of WVR (\ref{WittenVenez}) on YM topological susceptibility 
$\chi_{\rm Y\!M}$ implies $T$-dependence of $\eta'$ mass in conflict with 
the recent experimental results \cite{Csorgo:2009pa}.
Namely,
WVR is very remarkable because it connects two different theories:
QCD with quarks and its pure gauge, YM counterpart. The latter, however,
has much higher characteristic temperatures than QCD with quarks: the
``melting temperature" $T_{\rm Y\!M}$ where $\chi_{\rm Y\!M}(T)$
starts to decrease appreciably was found on lattice to be, for example,
$T_{\rm Y\!M} \approx 260$ MeV \cite{Alles:1996nm,Boyd:1996bx} or
even higher, $T_{\rm Y\!M} \approx 300$ MeV \cite{Gattringer:2002mr}.
In contrast, the pseudocritical temperatures for the chiral and
deconfinement transitions in the full QCD are lower than $T_{\rm Y\!M}$
by some 100 MeV or more (e.g., see Ref. \cite{Fodor:2009ax})
due to the presence of the quark degrees of freedom.

This difference in characteristic temperatures, in conjunction with
$\chi_{\rm Y\!M}(T)$ in WVR (\ref{WittenVenez})
would imply that the (partial) restoration of the U$_A$(1) symmetry
(understood as the disappearance of the anomalous $\eta_0/\eta'$ mass)
should happen well after the restoration of the chiral symmetry.
But, this contradicts the RHIC experimental observations of the
reduced $\eta'$ mass \cite{Csorgo:2009pa} {\it if}
WVR (\ref{WittenVenez}) holds unchanged also close to
the QCD chiral restoration temperature $T_{\rm Ch}$, around which
$f_\pi(T)$ decreases still relatively steeply
\cite{Horvatic:2007qs} for realistic explicit ChSB, thus leading 
to the increase of $6 \chi_{\rm Y\!M}(T)/f_\pi(T)^2$ and
consequently also of $M_{\eta'}$.

There is still more to the relatively high resistance of
$\chi_{\rm Y\!M}(T)$ to temperature: not only does it start falling
at rather high $T_{\rm Y\!M}$, but $\chi_{\rm Y\!M}(T)$
found on the lattice is falling with $T$ {\it relatively} slowly.
In some of the applications in the past (e.g., see Refs.
\cite{Fukushima:2001hr,Schaffner-Bielich:1999uj}), it was
customary to simply rescale a temperature characterizing the
pure gauge, YM sector to a value characterizing QCD
with quarks. (For example, Refs.
\cite{Fukushima:2001hr,Schaffner-Bielich:1999uj} rescaled
$T_{\rm Y\!M} = 260$ MeV found by Ref. \cite{Alles:1996nm} to 150 MeV).
However, even if we rescale the critical temperature for melting
of the topological susceptibility $\chi_{\rm Y\!M}(T)$ from
$T_{\rm Y\!M}$ down to $T_{\rm Ch}$, the value of $6 \chi_{\rm Y\!M}(T)/f_\pi(T)^2$
still increases a lot \cite{Horvatic:2007qs} for
the pertinent temperature interval starting already below $T_{\rm Ch}$.
This happens because $\chi_{\rm Y\!M}(T)$ falls with $T$ more slowly
than $f_\pi(T)^2$. 


One must therefore conclude that either WVR breaks down
as soon as $T$ approaches $T_{\rm Ch}$, or that the $T$-dependence
of its anomalous contribution is different from the pure-gauge
$\chi_{\rm Y\!M}(T)$.
We will show that the latter alternative is possible, since WVR
can be reconciled with experiment thanks to the existence of
another relation which, similarly to WVR, connects the YM theory
with full QCD.  Namely, using large-$N_c$ arguments,
Leutwyler and Smilga derived \cite{Leutwyler:1992yt}, at $T=0$,
\begin{equation}
\chi_{\rm Y\!M}\, = \,  \chi \,
\left( 1 + \chi\, \frac{N_f}{m\,\langle{\bar q}q\rangle_0} \right)^{-1}
\, \,  \left( \, \equiv \, {\widetilde \chi} \, \right)
 \, ,
\label{chitilde}
\end{equation}
the relation (in our notation) between the YM topological susceptibility
$\chi_{\rm Y\!M}$, and the full-QCD topological susceptibility $\chi$,
the {\it chiral-limit} quark condensate
$\langle{\bar q}q\rangle_0$, and $m$, the harmonic average of $N_f$
current quark masses $m_q$. That is, $m$ is $N_f$ times the reduced mass.
In the present case of $N_f=3$, $q=u,d,s$, so that
${N_f}/{m} = 1/m_u + 1/m_d + 1/m_s$

Eq. (\ref{chitilde}) is a remarkable relation between the two pertinent theories.
For example, in the limit of all very heavy quarks ($m_q\to \infty,\, q=u,d,s $),
it correctly leads to the result that $\chi_{\rm Y\!M}$ is equal to the value
of the topological susceptibility in {\it quenched} QCD,
$\chi_{\rm Y\!M} = \chi(m_q=\infty)$. This holds because $\chi$
is by definition the vacuum expectation value of a gluonic operator, so that
the absence of quark loops would leave only the pure-gauge, YM contribution.
However, the Leutwyler-Smilga relation (\ref{chitilde}) also holds
in the opposite (and presently pertinent) limit of light quarks.
This limit still presents a problem for getting the full-QCD
topological susceptibility $\chi$ on the lattice \cite{DeGrand:2007tx},
but we can use the light-quark-sector
result \cite{Di Vecchia:1980ve,Leutwyler:1992yt}
\begin{equation}
\chi = - \frac{m\,\langle{\bar q}q\rangle_0}{N_f} + {\cal C}_m
 \, ,
\label{chi_small_m}
\end{equation}
where ${\cal C}_m$ stands for corrections of higher orders in
small $m_q$, and thus of small magnitude.  The leading term
is positive (as $\langle{\bar q}q\rangle_0 <0$), but
${\cal C}_m$ is negative, since Eq. (\ref{chitilde}) shows that
$\chi \leq {\rm min}(- m\,\langle{\bar q}q\rangle_0/N_f, \chi_{\rm Y\!M})$.

Although small, ${\cal C}_m$ should not be neglected, since
${\cal C}_m = 0$ would imply $\chi_{\rm Y\!M} = \infty$, 
by Eq. (\ref{chitilde}).
Instead, its value (at $T=0$) is fixed by Eq. (\ref{chitilde}):
\begin{equation}
{\cal C}_m = {\cal C}_m(0) = \frac{m\,\langle{\bar q}q\rangle_0}{N_f}\, \left(
1 - \chi_{\rm Y\!M}\, \frac{N_f}{m\,\langle{\bar q}q\rangle_0}\right)^{-1} \, .
\label{Cat0}
\end{equation}

All this starting from Eq. (\ref{chitilde}) has so far been at $T=0$.
If the left- and right-hand side (RHS) of Eq. (\ref{chitilde})
are extended to $T>0$, it is obvious that the equality cannot hold
at arbitrary temperature $T>0$. The relation (\ref{chitilde}) must
break down somewhere close to the (pseudo)critical temperatures
of full QCD ($\sim T_{\rm Ch}$) since the pure-gauge quantity
$\chi_{\rm Y\!M}$ is much more temperature-resistant than RHS,
abbreviated as ${\widetilde \chi}$. The quantity ${\widetilde \chi}$,
which may be called the effective susceptibility, consists of the
full-QCD quantities $\chi$ and $\langle{\bar q}q\rangle_0$,
the quantities of full QCD with quarks, characterized by $T_{\rm Ch}$, 
just as $f_\pi(T)$.  As $T\to T_{\rm Ch}$, the chiral quark condensate 
$\langle{\bar q}q\rangle_0(T)$ drops faster than the other DChSB 
parameter in the present problem, namely $f_\pi(T)$ for realistically 
small explicit ChSB. (See Fig. 1 in our Ref. \cite{Benic:2011fv} 
for the results of the dynamical model adopted here from Ref.  
\cite{Horvatic:2007qs}, and, e.g., Refs. \cite{Maris:2000ig,Roberts:2000aa} 
for analogous results of different DS models).
Thus, the troublesome mismatch in $T$-dependences of $f_\pi(T)$
and the pure-gauge quantity $\chi_{\rm Y\!M}(T)$, which causes
the conflict of the temperature-extended WVR with experiment
around $T \geq T_{\rm Ch}$, is expected to disappear if
$\chi_{\rm Y\!M}(T)$ is replaced by  ${\widetilde \chi(T)}$,
the temperature-extended effective susceptibility.
The successful zero-temperature WVR (\ref{WittenVenez}) is, however,
retained, since $\chi_{\rm Y\!M} = {\widetilde \chi}$ at $T=0$.

Extending Eq. (\ref{chi_small_m}) to $T>0$ is something of a
guesswork as there is no guidance from the lattice for $\chi(T)$
[unlike $\chi_{\rm Y\!M}(T)$]. Admittedly, the leading term is
straightforward as it is plausible that its $T$-dependence will
simply be that of $\langle{\bar q}q\rangle_0(T)$. Nevertheless,
for the correction term ${\cal C}_m$ such a plausible assumption
about the form of $T$-dependence cannot be made and Eq. (\ref{Cat0}),
which relates YM and QCD quantities, only gives its value at $T=0$.
We will therefore explore the $T$-dependence of the anomalous masses
using the following Ansatz for the $T\geq 0$ generalization of
Eq. (\ref{chi_small_m}):
\begin{equation}
\chi(T) = - \frac{m\,\langle{\bar q}q\rangle_0(T)}{N_f} + {\cal C}_m(0) \,
\left[\frac{\langle{\bar q}q\rangle_0(T)}{\langle{\bar q}q\rangle_0(T=0)}\right]^{\delta}
 \, ,
\label{chi_small_mT}
\end{equation}
where the correction-term $T$-dependence is parametrized through the power
$\delta$ of the presently fastest-vanishing (as $T \to T_{\rm Ch}$)
chiral order parameter $\langle{\bar q}q\rangle_0(T)$.
In Eq. (\ref{chitildeChiCond}) below, it will become clear that
${\widetilde \chi}(T)$ blows up as $T \to T_{\rm Ch}$ if the correction
term there vanishes faster than $\langle{\bar q}q\rangle_0(T)$
{\it squared}.
Thus, varying $\delta$
between 0 and 2 covers the cases from the $T$-independent correction term,
to (already experimentally excluded) enhanced anomalous masses for $\delta$
noticeably above 1, to even sharper mass blow-ups for $\delta\to 2$
when $T \to T_{\rm Ch}$.
On the other hand, it is not natural that the correction term vanishes faster 
than the fastest-vanishing order parameter $\langle{\bar q}q\rangle_0(T)$.
This is why we depicted in Ref. \cite{Benic:2011fv} (in its Fig. 2) the 
$\delta=1$ case, and the $\delta=0$ ($T$-independent correction term) case, 
as two acceptable extremes. 
Since they turn out quite similar \cite{Benic:2011fv,Benic:2012eu} both 
qualitatively and quantitatively, there was no need to present any 
`in-between results', for $0 <\delta < 1$.


To see how the above-mentioned results were obtained, note that 
Eq. (\ref{chi_small_mT}) leads to the $T\geq 0$ extension
of ${\widetilde \chi}$ defined by Eq. (\ref{chitilde}):

\begin{equation}
\nonumber
{\widetilde \chi}(T) = \frac{m\,\langle{\bar q}q\rangle_0(T)}{N_f}
\left(1 - \frac{m\,\langle{\bar q}q\rangle_0(T)}{N_f \, {\cal C}_m(0)}
\left[\frac{\langle{\bar q}q\rangle_0(T=0)}{\langle{\bar q}q\rangle_0(T)}\right]^{\delta}
\right).
\label{chitildeChiCond}
\end{equation}
We now use ${\widetilde \chi}(T)$ in WVR instead of $\chi_{\rm Y\!M}(T)$
used in Ref. \cite{Horvatic:2007qs}.  Of course,
at $T=0$, ${\widetilde \chi}(T) = \chi_{\rm Y\!M}(0)$,
which remains an excellent approximation even well beyond $T=0$.
Nevertheless, this changes drastically as $T$ approaches $T_{\rm Ch}$.
For $T \sim T_{\rm Ch}$, the behavior of ${\widetilde \chi}(T)$
is dominated by the $T$-dependence of the chiral condensate,
tying the restoration of the U$_A$(1) symmetry to the
chiral symmetry restoration.

As for the non-anomalous contributions to the meson masses,
we use the same DS model
(and parameter values) as in Ref. \cite{Horvatic:2007qs},
since it includes both DChSB and correct QCD chiral behavior
as well as realistic explicit ChSB.
That is, all non-anomalous results
($M_\pi, f_\pi, M_K, f_K$, the chiral quark condensate
$\langle{\bar q}q\rangle_0$, as well as $M_{s\bar s}$ and
$f_{s\bar s}$, the mass and the decay constant of the unphysical
$s\bar s$ pseudoscalar meson, and $T$-dependences thereof) 
in the present paper are, for all
$T$, calculated in the model of Ref. \cite{Horvatic:2007qs}.  
For details of the non-anomalous sector, see Ref. 
\cite{Horvatic:2007qs}, see Refs. \cite{Kekez:2005ie,Klabucar:1997zi} for 
details on the construction of the $\eta_0$-$\eta_8$ complex,
Ref. \cite{Benic:2011fv} for the original paper on the present
topic, and Refs. \cite{Benic:2011fv,Benic:2012eu} for the 
detailed presentations of results.

\section{Conclusion}
\label{conclusion}

Thanks to the Leutwyler-Smilga relation (\ref{chitilde}),
the (partial) restoration
of U$_A$(1) symmetry [i.e., the disappearing contribution
of the gluon anomaly to the $\eta'$ ($\eta_0$) mass] is naturally
tied to the restoration of the SU$_A$(3) flavor chiral symmetry
and to its characteristic temperature $T_{\rm Ch}$, instead of
$T_{\rm YM}$.

In the both cases considered for the $T$-dependence of the
topological susceptibility (\ref{chi_small_mT})
[$\delta=0$, i.e., the constant correction term, and $\delta=1$, i.e.,
the strong $T$-dependence $\propto \langle{\bar q}q\rangle_0(T)$
of both the leading and correction terms in $\chi(T)$],
we find \cite{Benic:2011fv,Benic:2012eu} that $\eta'$ mass close 
to $T_{\rm Ch}$ suffers the drop of more than 200 MeV with respect
to its vacuum value. This satisfies the minimal experimental
requirement abundantly.
That is, the results are consistent with the experimental findings
on the decrease of the $\eta'$ mass of
Cs\"org\H{o} {\it et al.} \cite{Csorgo:2009pa}, 
as announced in the end of the Introduction.

We also note that our proposed mechanism, tying
${\widetilde \chi}$ to the chiral condensate
$\langle{\bar q}q\rangle_0$, suggests that partial 
$U_A(1)$-symmetry restoration would also happen if,
instead of temperature, matter density is increased 
sufficiently, so that the chiral symmetry restoration 
takes place and $\langle{\bar q}q\rangle_0$ vanishes. 


\vskip 3mm

 The support of the CompStar research
network, and the projects No. 0119-0982930-1016 and
No. 098-0982887-2872 of Ministry of Science, Education 
and Sports of Croatia, is acknowledged.


\begin{thebibliography}{10}


\bibitem{Csorgo:2009pa}
  T.~Csorgo, R.~Vertesi and J.~Sziklai,
  Phys.\ Rev.\ Lett.\  {\bf 105}, 182301 (2010).


\bibitem{Adler:2004rq}
  S.~S.~Adler {\it et al.}  [PHENIX Collaboration],
  Phys.\ Rev.\ Lett.\  {\bf 93}, 152302 (2004).


\bibitem{Adams:2004yc}
  J.~Adams {\it et al.}  [STAR Collaboration],
  Phys.\ Rev.\  C {\bf 71}, 044906 (2005).


\bibitem{Kapusta:1995ww}
  J.~I.~Kapusta, D.~Kharzeev and L.~D.~McLerran,
  Phys.\ Rev.\  D {\bf 53}, 5028 (1996).


\bibitem{Witten:1979vv}
E.~Witten,
\newblock Nucl. Phys. {\bf B156}, 269 (1979).

\bibitem{Veneziano:1979ec}
G.~Veneziano,
\newblock Nucl. Phys. {\bf B159}, 213 (1979).


\bibitem{Lucini:2004yh}
B.~Lucini, M.~Teper and U.~Wenger,
\newblock Nucl. Phys. {\bf B715}, 461 (2005).

\bibitem{DelDebbio:2004ns}
L.~Del~Debbio, L.~Giusti and C.~Pica,
\newblock Phys. Rev. Lett. {\bf 94}, 032003 (2005).

\bibitem{Alles:2004vi}
B.~Alles, M.~D'Elia and A.~Di~Giacomo,
\newblock Phys. Rev. {\bf D71}, 034503 (2005).

\bibitem{Durr:2006ky}
  S.~Durr, Z.~Fodor, C.~Hoelbling and T.~Kurth,
  JHEP {\bf 0704}, 055 (2007).


\bibitem{Horvatic:2007qs}
  D.~Horvati\'c, D.~Klabu\v{c}ar and A.~E.~Radzhabov,
  Phys.\ Rev.\  D {\bf 76}, 096009 (2007)
  [arXiv:0708.1260 [hep-ph]].


\bibitem{Benic:2011fv}
  S.~Beni\'c, D.~Horvati\'c, D.~Kekez and D.~Klabu\v{c}ar,
  Phys.\ Rev.\ D {\bf 84}, 016006 (2011)
  [arXiv:1105.0356 [hep-ph]].


\bibitem{Benic:2012eu}
  S.~Beni\'c, D.~Horvati\'c, D.~Kekez and D.~Klabu\v{c}ar,
  Acta Phys.\ Polon.\ Supp.\  {\bf 5}, 941 (2012)
  [arXiv:1207.3068 [hep-ph]].


\bibitem{Alles:1996nm}
B.~Alles, M.~D'Elia and A.~Di~Giacomo,
\newblock Nucl. Phys. {\bf B494}, 281 (1997).


\bibitem{Boyd:1996bx}
G.~Boyd {\em et~al.},
\newblock Nucl. Phys. {\bf B469}, 419 (1996), [hep-lat/9602007].


\bibitem{Gattringer:2002mr}
C.~Gattringer, R.~Hoffmann and S.~Schaefer,
\newblock Phys. Lett. {\bf B535}, 358 (2002). 


\bibitem{Fodor:2009ax}
  Z.~Fodor and S.~D.~Katz,
  arXiv:0908.3341 [hep-ph].


\bibitem{Fukushima:2001hr}
K.~Fukushima, K.~Ohnishi and K.~Ohta,
\newblock Phys. Rev. {\bf C63}, 045203 (2001).


\bibitem{Schaffner-Bielich:1999uj}
J.~Schaffner-Bielich,
\newblock Phys. Rev. Lett. {\bf 84}, 3261 (2000), [hep-ph/9906361].


\bibitem{Leutwyler:1992yt}
  H.~Leutwyler and A.~V.~Smilga,
  Phys.\ Rev.\  D {\bf 46}, 5607 (1992).


\bibitem{DeGrand:2007tx}
  T.~DeGrand and S.~Schaefer,
  arXiv:0712.2914 [hep-lat].


\bibitem{Di Vecchia:1980ve}
  P.~Di Vecchia and G.~Veneziano,
  Nucl.\ Phys.\  {\bf B171}, 253 (1980).


\bibitem{Maris:2000ig}
  P.~Maris {\it et al.},
  Phys.\ Rev.\  C {\bf 63}, 025202 (2001)
  [arXiv:nucl-th/0001064].


\bibitem{Roberts:2000aa}
  C.~D.~Roberts and S.~M.~Schmidt,
  Prog.\ Part.\ Nucl.\ Phys.\  {\bf 45}, S1 (2000).


\bibitem{Kekez:2005ie}
D.~Kekez and D.~Klabu\v{c}ar,
\newblock Phys. Rev. {\bf D73}, 036002 (2006) [hep-ph/0512064].

\bibitem{Klabucar:1997zi} 
  D.~Klabu\v{c}ar and D.~Kekez,
  Phys.\ Rev. {\bf D58}, 096003 (1998)
  [hep-ph/9710206].


\end{thebibliography}
\end{document}